 \documentclass[preprint2]{aastex631}

\def\dsm{$\mathrm{M}_\odot$}
\def\dsr{$R_\odot$}

\def\dalpha{$\alpha_{\rm{MLT}}$}
\def\dtcz{$T_{\mathrm{bcz}}$}

\shorttitle{Using Li and Be to Study Structure and Evolution of Rotating Stars}
\shortauthors{Yang et al.}

\begin{document}


\title{Using Lithium and Beryllium to Study Structure and Evolution of Rotating Stars}
\author[0000-0002-3956-8061]{Wuming Yang}
\affiliation{School of Physics and Astronomy, Beijing Normal University, Beijing 100875, China. }
\email{yangwuming@bnu.edu.cn}

\author{Haibo Yuan}
\affiliation{School of Physics and Astronomy, Beijing Normal University, Beijing 100875, China. }
\affiliation{Institute for Frontiers in Astronomy and Astrophysics, Beijing Normal University, Beijing, China.}
\author{Yaqian Wu}
\affiliation{Key Laboratory of Optical Astronomy, National Astronomical Observatories,
Chinese Academy of Sciences, A20 Datun Road, Chaoyang District, Beijing, 100101, China.}

\author{Shaolan Bi}
\affiliation{School of Physics and Astronomy, Beijing Normal University, Beijing 100875, China. }
\affiliation{Institute for Frontiers in Astronomy and Astrophysics, Beijing Normal University, Beijing, China.}

\author[0000-0003-0220-7112]{Zhijia Tian}
\affiliation{Department of Astronomy, Key Laboratory of Astroparticle Physics of Yunnan Province,
Yunnan University, Kunming 650200, China.}

\begin{abstract}
The chemical composition of the Sun is still a highly controversial issue. No solar model has
yet been able to simultaneously reproduce the solar lithium and beryllium abundances, along with
helioseismic results, including the rotation profile. Lithium and beryllium are fragile elements that
are highly sensitive to the physical conditions, as well as to transport and mixing processes within
and below the convective zone (CZ). Uncovering the transport mechanisms responsible for the depletion
of Li and Be in the Sun is crucial for using them as tools to understand stellar interiors
and the associated transport and mixing processes. We constructed rotating solar models based on
Magg's abundance scale, incorporating the effects of convective overshoot and magnetic fields.
The rotating model exhibits superior sound speed and density profile and successfully reproduces
the observed ratios $r_{02}$ and $r_{13}$. It also matches the seismically inferred CZ depth, surface
helium abundance, and rotation profile, as well as the detected Li and Be abundances and neutrino
fluxes within $1\sigma$. The depletion of Li is dominated by convective overshoot and rotational mixing,
while Be depletion is primarily driven by gravitational settling and rotational mixing.
The presence of the tachocline accelerates Li depletion but slows down Be depletion. These
distinct depletion mechanisms result in the surface abundances of Li and Be evolving differently
over time.
\end{abstract}


\section{INTRODUCTION}

Recent analyses infer that lithium (Li) abundance, $A$(Li), in the solar photosphere and thus
the convection zone (CZ) is 0.96 $\pm$ 0.05 dex \citep{wang21}, while beryllium (Be) abundance,
$A$(Be), is 1.32 $\pm$ 0.05 dex \citep{carlb18, koro22}, where $A(\rm x)=\log(N_{x}/N_{H})+12.0$.
However, at the time of the birth of the Sun, the Li and Be abundances were $A$(Li) = 3.3 dex,
as derived from meteorites \citep{lodd21}, and $A$(Be) = 1.44 dex, as determined from solar
observations with some corrections \citep{aspl21}, respectively.

Since \citet{green51} and \citet{green54} first measured the abundances of Li and Be in
the solar photosphere, astronomers have recognized that the solar Li is significantly depleted,
whereas Be almost remains unchanged (though see \citet{chmi75} and \citet{bala98} for further
discussions). This phenomenon is commonly referred to as solar lithium depletion.

The elements $^{7}$Li and $^{9}$Be are easily destroyed by energetic proton at temperatures near $2.5\times10^{6}$ K
and $3.5\times10^{6}$ K, respectively, making them fragile elements. Thus,$^{7}$Li and $^{9}$Be
primarily survive in the outer regions of stars with temperature $T\lesssim2.3\times10^{6}$ K and $T\lesssim3.0\times10^{6}$
K, respectively. The depth of the Sun's CZ can be measured using helioseismology. The inferred radius of
the base of the CZ (BCZ), $r_{\rm cz}$, is 0.713 $\pm$ 0.003\dsr{} \citep{chri91} or 0.713 $\pm$ 0.001\dsr{}
\citep{basu97}. The Sun's rotation profile can also be determined via helioseismology \citep{thom03, eff08}.
Solar models predict that the temperature at the BCZ is about $(2.20\pm0.04)\times10^{6}$ K \citep{yang24}.
Therefore, the convection in the present Sun cannot account for the depletion
of Li and Be in the CZ. This indicates that a mixing mechanism must be operating in the
solar radiative region, especially below the CZ, to transport Li and Be from the BCZ
into a higher-temperature region where they are burned. Thus, Li and Be abundances provide valuable
insights into the interiors of stars and physical processes occurring within them. Both Li and Be are excellent
astrophysical tracers of transport and mixing processes within and below the CZ of solar-type stars.

To better understand stellar interiors and the mechanisms behind Li and Be depletion, the
abundances of these elements have been measured in many solar twins and solar-type stars \citep{smil11, take11, galv11,
carl19, carl20, boes20, boes22}. The Li abundances of other late-type stars have also been determined
\citep{ding24, wang24}. With $A$(Li)$_{\odot}=1.07^{+0.03}_{-0.02}$ dex, \citet{carl19} found that
the Sun has the lowest Li abundance compared to solar twins at a similar age. \citet{boes20, boes22}
obtained similar results, showing that Sun's A(Li) falls within the lower third of
their total sample of solar-mass stars, whereas A(Be) is in the upper third. This suggests that
Li and Be depletion may arise from different physical mechanisms. The Sun appears to be
exceptionally depleted in Li but only slightly depleted in Be. These characteristics provide strong
constraints on the transport and mixing mechanisms below the CZ, as well as their efficiencies.

Many stellar models incorporating different physics have been used to explain the Li depletion
of the Sun and solar-like stars. For example, \citet{charb05} considered the effects of gravity waves;
\citet{xiong09} accounted for overshooting and gravitational settling; while rotation and diffusion
were included in the models of \citet{dona09}, \citet{dumo21a, dumo21b}, \citet{cons21}, \citet{egge22}, and \citet{yang24}.
The solar Li abundance alone can be easily reproduced by an evolutionary model. However, simultaneously
matching the observed Li, Be, and other solar characteristics remains challenging. For instance, the rotating
model of \citet{egge22} successfully reproduces the He and Li abundances, as well as the rotation profile of the Sun,
but fails to match the observed Be abundance, the seismically inferred CZ depth, and the sound speed profile.
This highlights the complexity of solar modeling.

Moreover, the chemical composition of the Sun remains a highly debated issue \citep[and references therein]
{aspl21, magg22, piet23, yang24}. To replicate the solar structure and element abundances, the effects of various
accretion processes have been considered \citep{zhang19, kuni22}. Rotational mixing has been proposed as a
solution to the low helium abundance problem in solar models based on recently estimated solar
metallicity \citep{yang07, yang16, yang19, yang22, yang24}. However, there is no direct evidence that
rotational mixing increases the Sun’s surface helium abundance. Mechanisms that transport material into the
Sun's radiative region would also carry helium-rich material from the radiative region to the CZ,
potentially enhancing the surface helium abundance. If these mechanisms are driven by rotation, then
a rotating solar model should be able to simultaneously reproduce the observed surface Li and Be abundances,
the seismically inferred surface helium abundance, CZ depth, and rotation profile, as well as
the surface rotation history speculated from solar-type main-sequence (MS) stars in clusters. In this scenario,
the depletion of Li and Be would serve as evidence that rotation enhances surface helium abundance. Thus, obtaining
an accurate solar model is crucial for addressing the solar composition problem (or solar modeling problem).

The entire problem of the abundances of Li and Be in the Sun, which has puzzled astronomers for over six decades, requires
further close scrutiny. It is imperative to uncover transport mechanisms responsible for the depletion of Li and Be in the Sun
in order to use these elements as tracers for understanding stellar interiors and the processes of transport and mixing
within them. Furthermore, a full understanding of Li depletion also aids in accurately determining
the primordial Li abundance from observations of the Galaxy's oldest metal-poor stars.
In this study, we present rotating solar models that can simultaneously reproduce the observed
Li and Be abundances of the Sun, the seismically inferred rotation profile, and other key characteristics.
Our findings demonstrate that the depletion of Li and Be is dominated by different mechanisms.
The paper is organized as follows. Input physics are presented in Section 2, calculation results are shown in Section 3,
and the results are discussed and summarized in Section 4.

\section{Input Physics}
Stellar evolutionary models were computed using the Yale Rotating Stellar Evolution Code \citep{enda76, pins89,
yang07, yang16} in its rotation and nonrotation configurations. The OPAL equation-of-state (EOS2005)
tables \citep{roge02} and the Opacity Project (OP) \citep{seat87, badn05} or OPAL \citep{igle96} opacity tables were utilized,
supplemented by the \citet{ferg05} opacity tables at a low temperature. These tables were reconstructed
with \citet{magg22} mixtures. The Rosseland mean opacity between $\log T=$ 4.1 and 4.0 was
determined through linear interpolation of the two opacity tables. The nuclear reaction rates were
calculated using the subroutine of \citet{bahc92}, updated by \citet{bahc95, bahc01} and \citet{yang24}.
The nuclear reaction rates of Li and Be as a function of temperature were calculated for temperatures
exceeding $10^{6}$ K, under the assumption that Li and Be are completely destroyed
at $T\geq10^{7}$ K. The diffusion and settling of both helium and heavy elements were computed using
the diffusion coefﬁcients of \citet{thou94}. The effects of radiative levitation on chemical transport
were not incorporated into our models. The radiative effects can cause the abundance and mixture of
the heavy elements to vary with stellar age and position in the Sun \citep{turc98}. However, the impact
of radiative acceleration may be mitigated by the effects of rotation and magnetic fields.

Li and Be abundances are generally not included in the metal abundance $Z$ because their concentrations are
too low. Similarly, the diffusion and settling of heavy elements do not account for Li and Be. However,
in this work, we incorporated the diffusion and settling of Li and Be into all models.

In the atmosphere, the \citet{kris66} $T-\tau$ relation was adopted. The boundary of the CZ was calculated
using the Schwarzschild criterion, and energy transfer by convection was treated according to the
standard mixing-length theory \citep{bohm58}. The depth of the overshoot region was given
by $\delta_{\rm ov}H_{p}$, where $\delta_{\rm ov}$ is a free parameter, and $H_{p}$ is the local
pressure scale height. The overshoot region was assumed to be both fully mixed and adiabatically stratified.

The angular momentum loss from the CZ due to magnetic braking was computed using Kawaler’s relation
\citep{kawa88, chab95}. The redistributions of angular momentum and chemical compositions were
treated as a diffusion process \citep{enda78, yang16}, i.e.,
    \begin{equation}
       \frac{\partial \Omega}{\partial t}=
       \frac{1}{\rho r^{4}}\frac{\partial}{\partial r}[(\rho r^{4}(f_{\Omega}D_{r}+f_{m}D_{m})
       \frac{\partial \Omega}{\partial r})] \,
      \label{diffu1}
    \end{equation}
for angular momentum transport and
    \begin{equation}
    \begin{array}{lll}
        \frac{\partial X_{i}}{\partial t}&=&\frac{1}{\rho r^{2}}
       \frac{\partial}{\partial r}[\rho r^{2}(f_{c}f_{\Omega}D_{r}+f_{cm}f_{m}D_{m})\frac{\partial X_{i}}
        {\partial r}]\\
        & &+(\frac{\partial X_{i}}{\partial t})_{\rm nuc}-\frac{1}
       {\rho r^{2}}\frac{\partial}{\partial r}(\rho r^{2}X_{i}V_{i}) \,
    \end{array}
      \label{diffu2}
    \end{equation}
for the change in the mass fraction $X_{i}$ of chemical species $i$, where $D_{r}$ is the diffusion
coefficient caused by rotational instabilities, including the dynamical instabilities described in
\citet{enda78} and \citet{pins89}, as well as the secular shear instability
\begin{equation}
 D = \frac{2c}{27G}|\frac{d\ln T}{dr}-\frac{2}{3}
       \frac{d\ln\rho}{dr}|^{-1}\frac{r^{4}}{\kappa \rho M(r)}
       (\frac{d\Omega}{dr})^{2}
\end{equation}
of \cite{zahn93}; $\rho$ is the density; and $V_{i}$ is the velocity of microscopic diffusion
and settling given by \citet{thou94}; $\kappa$ is the opacity; $c$ and $G$ are the light
velocity and gravitational constant, respectively. The diffusion coefficient due to magnetic
fields, $D_{m}$, is defined by \citep{yang06}
\begin{equation}
 D_{m}=r^{2}\Omega\frac{B^{2}_{r}}{B^{2}}.
\end{equation}
We use the radial ($B_{r}$) and toroidal ($B_{t}$) components to express a magnetic ﬁeld vector
$\mathbf{B}$ = $(B_{r}, B_{t})$. Given that $|B_{t}|\gg|B_{r}|$, here, we take $|B|\approx|B_{t}|$.
The magnetic ﬁeld compositions $|B_{r}|$ and $|B_{t}|$ are calculated using Equations (22)
and (23) of \citet{spru02}. The parameters $f_{\Omega}$ and $f_{m}$ were introduced to represent
some inherent uncertainties in the diffusion equation, while the parameters $f_{c}$ and $f_{cm}$
were used to account for how the instabilities and magnetic fields mix material less efficiently
than they transport angular momentum \citep{pins89, yang16}. The default values of $f_{\Omega}$
and $f_{c}$ are $1$ and $0.03$, respectively.

All solar models are calibrated to the present solar luminosity $3.844\times10^{33}$ erg s$^{-1}$,
radius $6.9598\times10^{10}$ cm, mass $1.9891\times10^{33}$ g, and age $4.57$ Gyr \citep{bahc95}.
The initial hydrogen abundance $X_{0}$, metal abundance $Z_{0}$, and mixing-length parameter \dalpha{}
are free parameters adjusted to match the constraints of luminosity and radius around $10^{-5}$
and an observed $(Z/X)_{s}$. The initial helium abundance is determined by $Y_{0}=1-X_{0}-Z_{0}$.
The initial rotation rate, $\Omega_{i}$, of rotating models is also a free parameter.

We constructed the following five models: (1) MBS22, a standard solar model (SSM) constructed using
OP opacity tables with \citet{magg22} mixture; (2) MBS22r1, a rotating model that includes the effects of
magnetic fields \citep{yang16} and is constructed using OPAL opacity tables with the \citet{magg22} mixture;
(3) MBS22r2 and MBS22r3, which are similar to MBS22r1 but have higher mixing efficiency; (4) MBS22rt,
similar to MBS22r2 but with an assumed tachocline of $0.05 R$ and a different mixing efficiency within the tachocline,
where $R$ represents the stellar radius. The four rotating models share the same $f_{m}$ but have
different $f_{cm}$. The fundamental parameters of these models are listed in Table \ref{tab1}.

\begin{deluxetable*}{lcccccccccccccc}
\tablecaption{Fundamental Parameters of Models.
\label{tab1}}
\tablewidth{0pt}
\tablehead{
 Model &  $Y_{0}$   & $Z_{0}$ & \dalpha{} & $\delta_{\rm ov}$ & $r_{\rm cz}$ & \dtcz{} & $Y_{s}$ & $Z_{s}$ & $(Z/X)_{s}$ &
$\Omega_{\rm z}$ & $f_{m}$ & $f_{cm}$ & $A$(Li)$_{s}$ & $A$(Be)$_{s}$   }
\startdata
 MBS22            & 0.27066  & 0.0183 & 2.0973 & 0    & 0.718 & 2.12 & 0.2406 & 0.01646 & 0.0222 & 0          &...&...& 2.45 & 1.30 \\
 MBS22r1 & 0.27392  & 0.0184 & 2.0755 & 0.09 & 0.711 & 2.21 & 0.2486 & 0.01638 & 0.0223 & $\sim$10   & 1 & 1 & 1.40 & 1.33 \\
 MBS22r2 & 0.27391  & 0.0184 & 2.0691 & 0.09 & 0.713 & 2.20 & 0.2500 & 0.01640 & 0.0223 & $\sim$10   & 1 & 2 & 1.25 & 1.32 \\
 MBS22r3 & 0.27391  & 0.0184 & 2.0556 & 0.09 & 0.713 & 2.20 & 0.2529 & 0.01641 & 0.0225 & $\sim$10   & 1 & 6 & 0.82 & 1.22 \\
 MBS22rt & 0.27391  & 0.0184 & 2.0613 & 0.09 & 0.713 & 2.20 & 0.2517 & 0.01641 & 0.0224 & $\sim$10   & 1 & 2$^{a}$ & 0.94 & 1.31 \\
 \enddata
\tablenotetext{}{Notes. The CZ radius $r_{\rm cz}$, CZ temperature \dtcz{}, angular velocity $\Omega_{\rm z}$, $f_{m}$, and $f_{cm}$
are in units of $R_{\odot}$, $10^{6}$ K, $10^{-6}$ rad s$^{-1}$, $10^{-4}$, and $10^{-4}$, respectively. $^{a}$ The value of $f_{cm}$
within the tachocline of MBS22rt is 0.0005.}
\end{deluxetable*}

\section{CALCULATION RESULTS}
\subsection{Results of Standard Solar Model}

The surface Li and Be abundances of MBS22 are $2.45$ and $1.30$, respectively. The predicted Be abundance
is in agreement with the observed value of $1.32\pm 0.05$, but the Li abundance is too high. Moreover, the CZ base
radius of $0.718$ \dsr{} and the surface helium abundance of $0.2406$ disagree with the seismically inferred
values. The position of the BCZ is too shallow, and the surface helium abundance is too low.

\begin{figure}
\includegraphics[angle=0, scale=0.6]{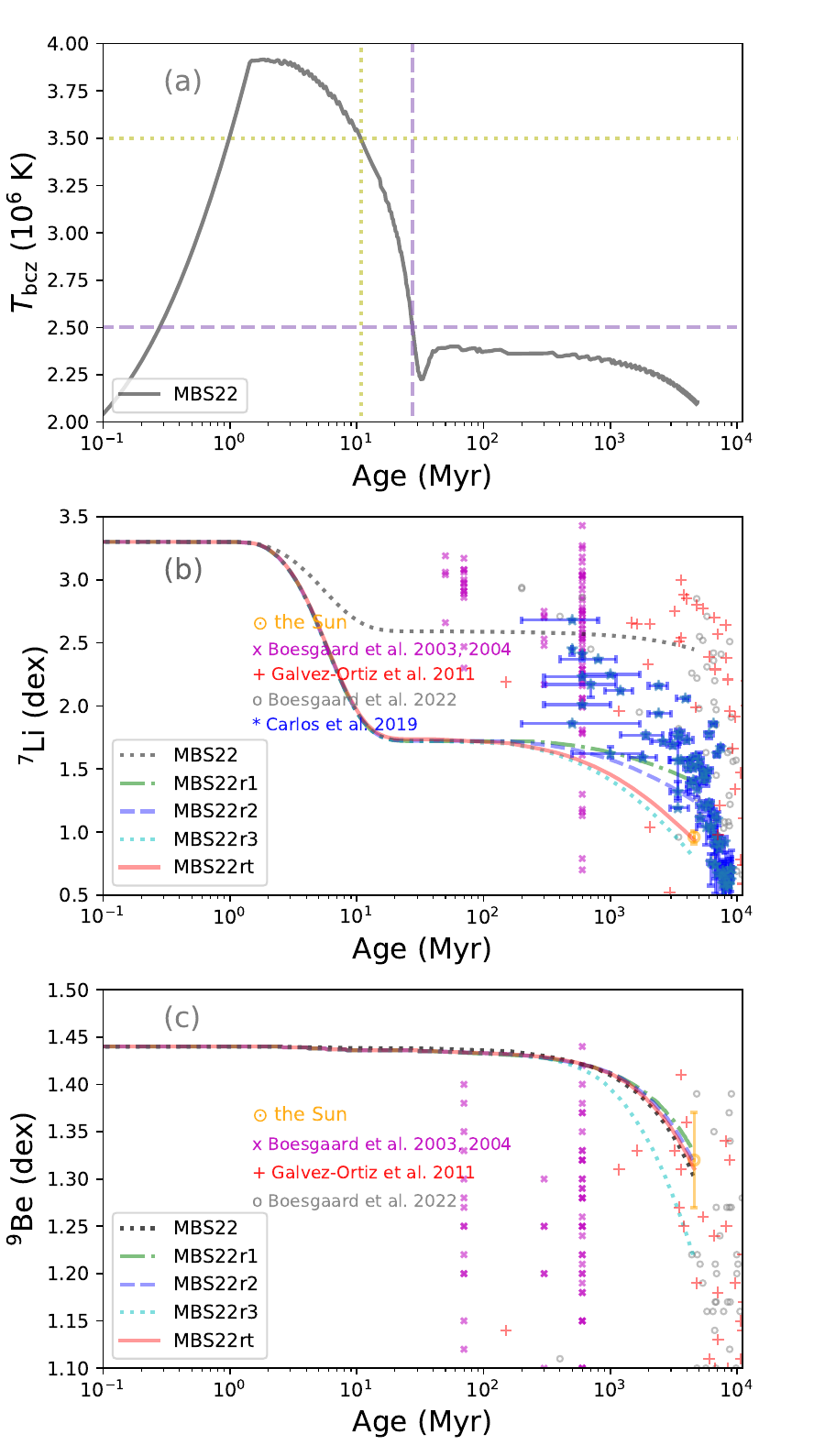}
\caption{(a) Temperature at the BCZ of nonrotating models as a function of age. The vertical
dotted and dashed lines represent the ages at \dtcz{} = $3.5\times10^{6}$ and $2.5\times10^{6}$ K, respectively.
(b), (c) Surface lithium and beryllium abundances as a function of age for different solar models. The age of the solar models
at the zero-age main sequence (ZAMS) is about $40$ Myr. The scatter symbols refer to estimated Li or Be abundance \citep{boes03a,
boes03b, boes04, boes22, galv11, carl19, wang21, koro22}.
\label{fig1}}
\end{figure}

\begin{figure*}
\includegraphics[angle=0, scale=0.45]{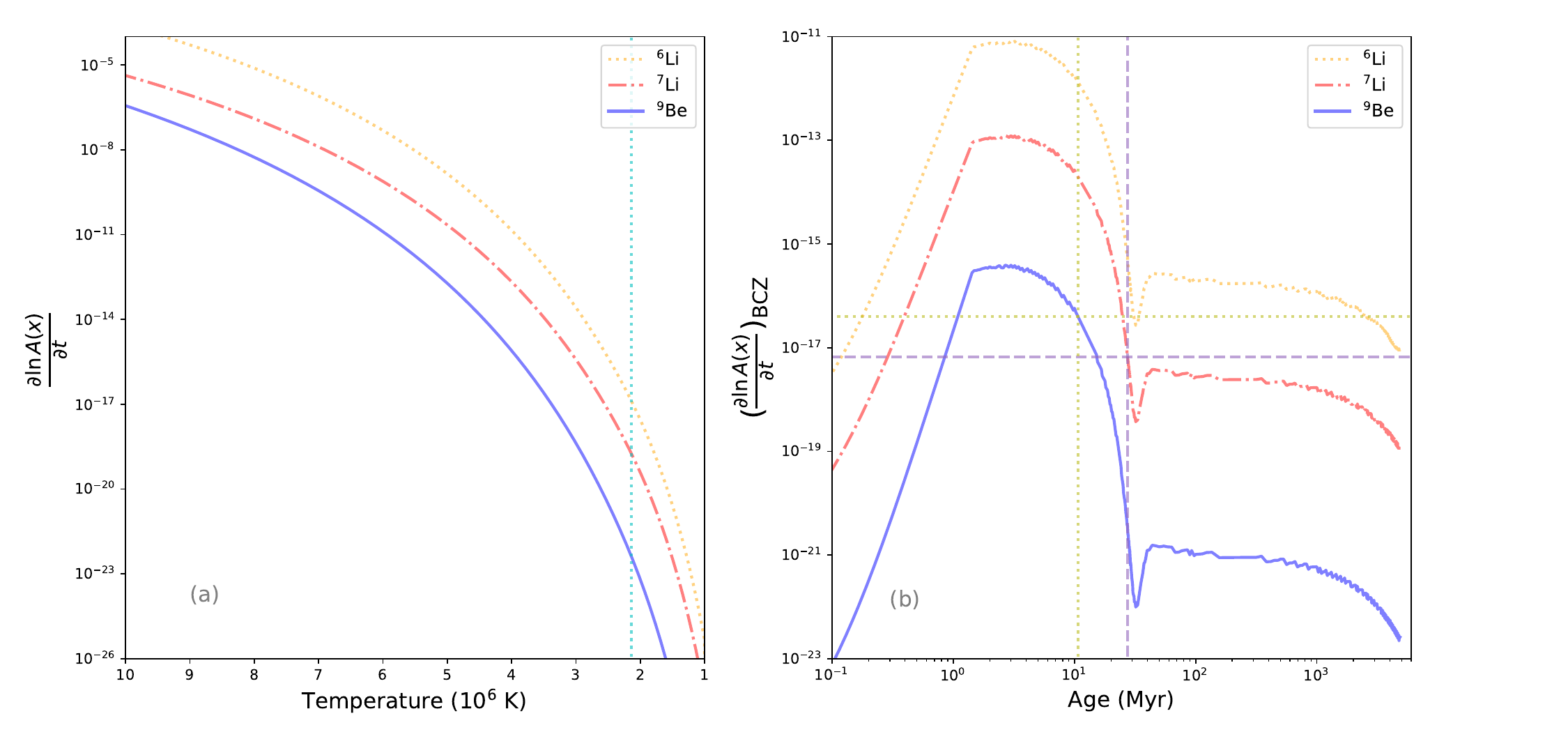}
\caption{(a) Burning rates of $^{6}$Li, $^{7}$Li, and $^{9}$Be of MBS22 as a function of temperature.
The vertical dotted line represents the BCZ temperature. (b) Burning rates of $^{6}$Li, $^{7}$Li, and $^{9}$Be
at BCZ as a function of age. The vertical dotted and dashed lines represent the ages at \dtcz{} =
$3.5\times10^{6}$ and $2.5\times10^{6}$ K, respectively.
\label{fig2}}
\end{figure*}

Figure \ref{fig1} shows that the temperature at the BCZ, \dtcz{}, exceeds $3.5\times10^{6}$ K during
the early pre-main-sequence (pre-MS) stage, which can destroy Be at the BCZ. However, this phase lasts for only about
$10$ Myr. The amount of Be burned at the BCZ during this stage is negligible. Nevertheless, the
temperature is high enough to significantly deplete Li at the BCZ due to a higher burning rate
(see Figure \ref{fig2}). Consequently, the surface Be abundance remains almost constant during the pre-MS stage,
while the surface Li abundance decreases with age due to the effects of convection (see Figure \ref{fig1}).

During the MS stage, the temperature \dtcz{} does not exceed $2.4\times 10^{6}$ K (see Figure \ref{fig1}).
Since the burning rates of $^{7}$Li and $^{9}$Be in the CZ of MBS22 are low and
decrease rapidly with a decrease in temperature ($\partial \ln A(\rm Li)/\partial t < 4\times10^{-18}$
s$^{-1}$ and $\partial \ln A(\rm Be)/ \partial t < 2\times10^{-21}$ s$^{-1}$; see Figure \ref{fig2}), their
burning in the CZ is insufficient to significantly alter their abundances.
The changes in the surface Li and Be abundances of MBS22 during this stage are primarily due to
gravitational settling rather than nuclear fusion reactions in the CZ. Gravitational settling
is the dominant mechanism affecting the surface Be abundance. To accurately understand the behavior of Li and Be,
the effects of gravitational settling must be considered. If gravitational settling were neglected, the surface Li and Be
abundances would remain unchanged during the MS stage of SSMs.

From MBS22, it is evident that the seismically inferred CZ depth, surface He abundance, and Li
abundance do not support the SSM. Additional mixing mechanisms, such as rotational effects, are
required to simultaneously explain the solar Li and Be depletion, as well as other observed
characteristics.

\subsection{Results of Rotating Models}

To reproduce the observed Li abundance and other characteristics, we considered the effects of rotation,
including magnetic fields \citep{yang16}, and constructed four rotating models with different
mixing efficiencies. In these models, convective overshooting with $\delta_{\rm ov}=0.09$ is necessary
to reproduce the seismically inferred CZ depth. Convective overshooting carries material from the CZ
into a region of higher temperature. During the pre-MS stage, convection and overshooting predominantly
influence the change in the Sun's surface Li abundance, as the temperature \dtcz{} during this stage is high
enough to rapidly destroy Li (see Figure \ref{fig1}). However, these processes have little effect
on the surface Be abundance and the Li abundance during the MS stage.

The effects of convective overshooting alone cannot explain solar Li depletion unless we ignore
the constraints from helioseismology, such as the seismically inferred CZ depth and the observed
frequency separation ratios $r_{02}$ and $r_{13}$ that are influenced by the acoustic depth of the
CZ \citep{roxb03}, and adopt a larger $\delta_{\rm ov}$. During the MS stage, changes in surface
Li and Be abundances in rotating models are driven by magnetic and hydrodynamic mixing,
alongside gravitational settling.

\begin{figure*}
\includegraphics[angle=0, scale=0.5]{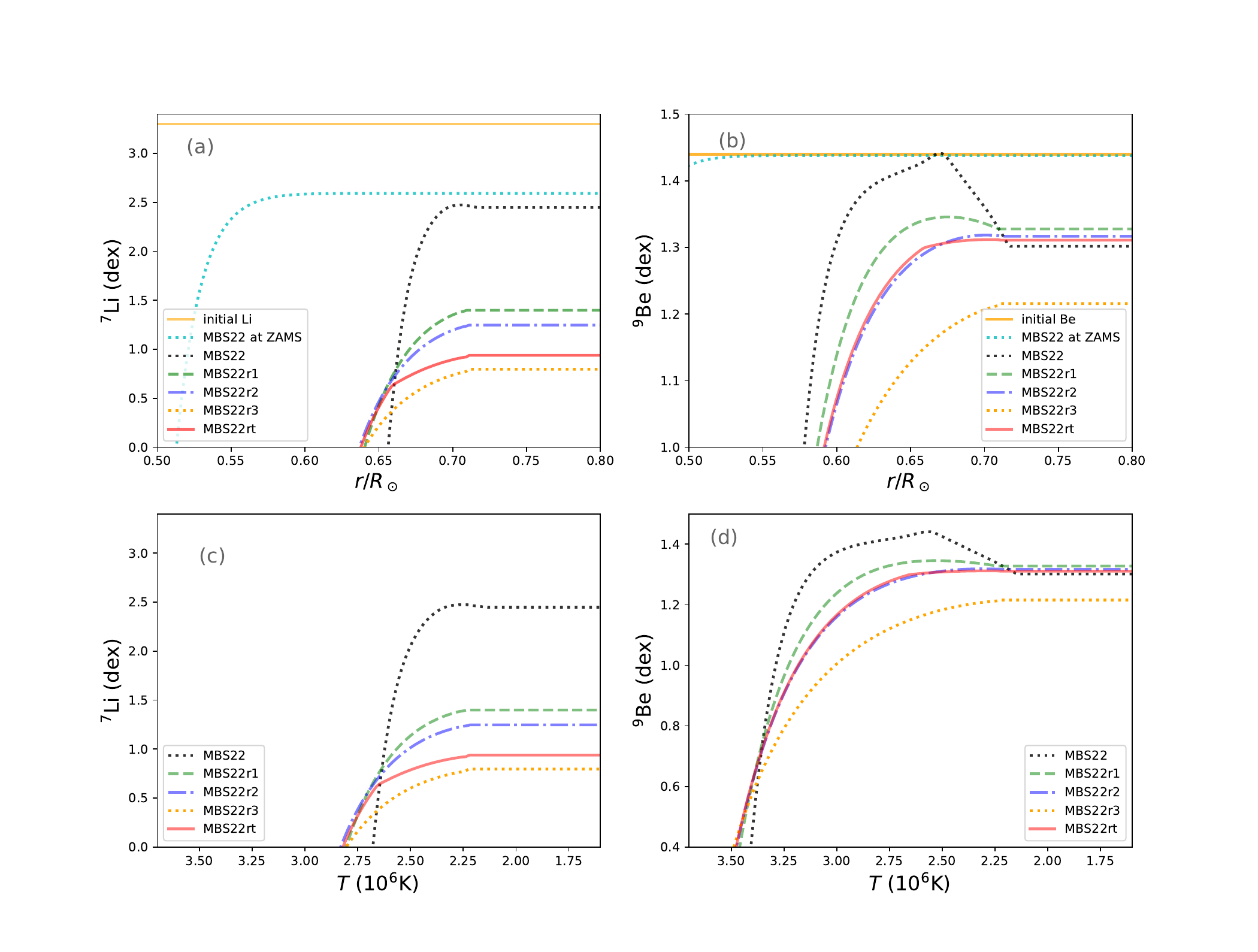}
\caption{(a), (b) Lithium and beryllium profiles as a function of radius for different models.
(c), (d) Lithium and beryllium profiles as a function of temperature for different models.
\label{fig3}}
\end{figure*}

In the nonrotating model, Li from the CZ that settles near the top of the radiative region does not
accumulate there because the temperature is high enough to destroy it, and diffusion removes it.
However, Be from the CZ can partially accumulate in that region since the temperature is not high
enough to effectively burn it, and diffusion only removes a fraction of the Be.
As a result, Li and Be exhibit different distribution profiles (see the black dotted lines in Figure \ref{fig3}).

In rotating models, rotational mixing transports Li and Be to regions with higher temperatures
(see Figure \ref{fig3}), where they are burned more rapidly. When Li from the CZ
is brought into the radiative region, helium from the radiative region
can be transported to the CZ. This process leads to an increase in surface helium abundance.
Therefore, in a rotating star, surface Li depletion and He enhancement occur simultaneously.
If the efficiency of rotational mixing is too high, it will result in surface Li and Be abundances
being too low, while the surface He abundance becomes excessively high.

With $f_{m}$ = 0.0001 and $f_{cm}$ = 0.0001, we constructed the rotating model MBS22r1. The surface
He, Li, and Be abundances predicted by MBS22r1 for the Sun are 0.2486, 1.40, and 1.33 dex,
respectively. The surface He and Be abundances are consistent with inferred values; however,
the surface Li abundance is too high.

\begin{figure*}
\includegraphics[angle=-90, scale=0.75]{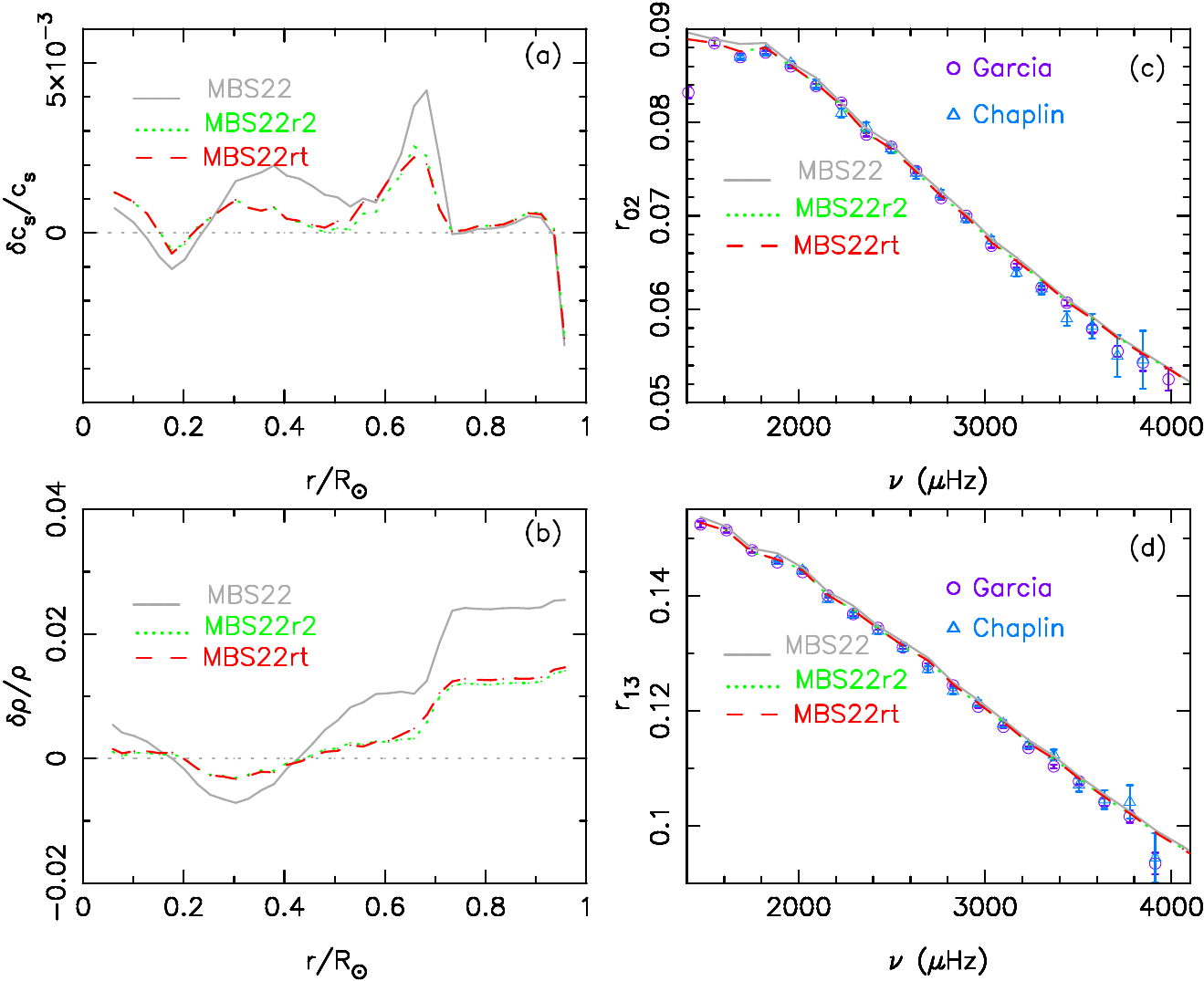}
\caption{(a), (b) Relative sound speed and density differences, in the sense (Sun-Model)/Model, between
the Sun and models. The inferred sound speed and density of the Sun are given by \citet{basu09}.
(c), (d) Distributions of observed and predicted ratios $r_{02}$ and $r_{13}$ as a function of frequency,
which are calculated by using smooth ﬁve-point separations of \citet{roxb03}.
The circles and triangles show the ratios calculated from the frequencies observed by GOLF and
VIRGO \citep{garc11} and by BiSON \citep{chap99}, respectively.
\label{fig4}}
\end{figure*}

With $f_{cm}$ = 0.0002, we constructed the rotating model MBS22r2. This model has better sound speed
and density proﬁles than SSM MBS22 and reproduces the observed ratios $r_{02}$ and $r_{13}$ (see Figure \ref{fig4}).
The surface helium abundance of 0.2500 and the CZ base radius of 0.713 \dsr{} of MBS22r2 are in good
agreement with the seismically inferred values. The surface Be abundance of 1.32 dex matches
the 1.32 $\pm$ 0.05 dex determined by \citet{koro22}. However, the surface Li abundance of 1.25 dex remains
higher than the 1.04 $\pm$ 0.10 dex advocated by \citet{lodd21} or the 0.96 dex reported by \citet{wang21}.

Models MBS22r1 and MBS22r2 show that surface He abundance increases while Li abundance decreases with
higher mixing efficiency. However, the surface Li abundances in both models remain higher than the observed
value, suggesting that the mixing efficiency may be underestimated. Thus, we constructed the rotating model
MBS22r3 with $f_{cm}$ = 0.0006. The surface He, Li, and Be abundances predicted by MBS22r3 for the Sun
are $0.2529$, $0.82$ dex, and $1.22$ dex, respectively. The surface He abundance is higher than the value inferred from
helioseismology, while the surface Li and Be abundances are lower than the observed values, indicating that
MBS22r3 likely overestimates the Sun's mixing efficiency.

Moreover, we also constructed a rotating model with $f_{cm}$ = 0.0004. The surface Li and Be abundances
predicted by this model for the Sun are 1.03 and 1.27 dex, respectively. When adopting the Li abundance of
$0.96\pm 0.05$ dex advocated by \citet{wang21} and the Be abundance of 1.32 dex to constrain solar models,
it becomes apparent that this model underestimates the mixing of Li while overestimating that of Be.
Helioseismology has shown that the Sun has a tachocline with a width of $0.039\pm0.013$ \dsr{} below the CZ
\citep{char99}, where latitudinal differential rotation occurs. The shear in the tachocline
is expected to be very strong; therefore, a component associated with the hydrodynamical transport of material
should be present. Consequently, it is widely believed that rotational mixing is more efficient in the tachocline
than in other regions.

Assuming that the width of the tachocline is $0.05R$ and the mixing is more efficient in tachocline than in
other regions, we achieved this by setting $f_{cm}$ = 0.0005 and $f_{c} = 0.03\times2.5$ in the tachocline,
while using $f_{cm}$ = 0.0002 and $f_{c} = 0.03$ in other regions. With these assumptions, we constructed
the model MBS22rt. The surface helium abundance of 0.2517 and the CZ base radius of 0.713 \dsr{}
for MBS22rt are in good agreement with the seismically inferred values. It also exhibits superior
sound speed and density proﬁles compared to MBS22 and reproduces the observed ratios $r_{02}$ and $r_{13}$
(see Figure \ref{fig4}). Additionally, the surface Li abundance of 0.94 dex in MBS22rt agrees with the
$0.96\pm 0.05$ dex advocated by \citet{wang21}, while the surface Be abundance of 1.31 dex
is consistent with the 1.32 $\pm$ 0.05 dex reported by \citet{koro22}.

Table \ref{tab1} shows that the surface Be abundances in MBS22r1, MBS22r2, and MBS22rt are higher,
whereas that in MBS22r3 is lower than the value predicted by MBS22. Panel (b) of Figure \ref{fig3}
illustrates that the Be abundance below the CZ in MBS22r1 exceeds that within the CZ. The temperature at the base
of tachocline of a solar model is about $2.6\times 10^{6}$ K, which is insufficient to effectively destroy beryllium.
Gravitational settling causes the Be abundance in the CZ to accumulate in the region below it, resulting in a lower
Be abundance within the CZ compared to the region beneath it. When the rotational mixing efficiency is low,
Be below the CZ cannot be efficiently transported to higher-temperature regions where it would be burned.
In such cases, the negative Be abundance gradient causes rotational mixing below the CZ to partially
counteract the effect of gravitational settling. Consequently, the surface Be abundance of a rotating
model with low mixing efficiency is higher than that of a nonrotating model.

However, when the mixing efficiency is high, as in MBS22r3, Be below the CZ is rapidly transported to
higher-temperature regions and burned. This results in a positive Be abundance gradient below the CZ (see Figure
\ref{fig3}), with rotational mixing transporting Be from the CZ to regions where it is destroyed. Consequently,
the surface Be abundance in a high-efficiency rotating model is lower than that in a nonrotating model.
The Sun's surface Be abundance is influenced by gravitational settling, rotational mixing, and the tachocline.
Unlike Be, the temperature in the tachocline is sufficient to effectively destroy Li, leading to different behavior
for these elements.

Our calculations show that the upper limit of the surface Be abundance predicted by rotating models
is $1.34$ dex, determined by the interaction between gravitational settling and rotational
mixing. Accurately and precisely determining the Sun's Be abundance is crucial for understanding
these processes and their interactions. Be serves as an excellent tracer of these physical
processes.

\begin{figure*}
\includegraphics[angle=0, scale=0.4]{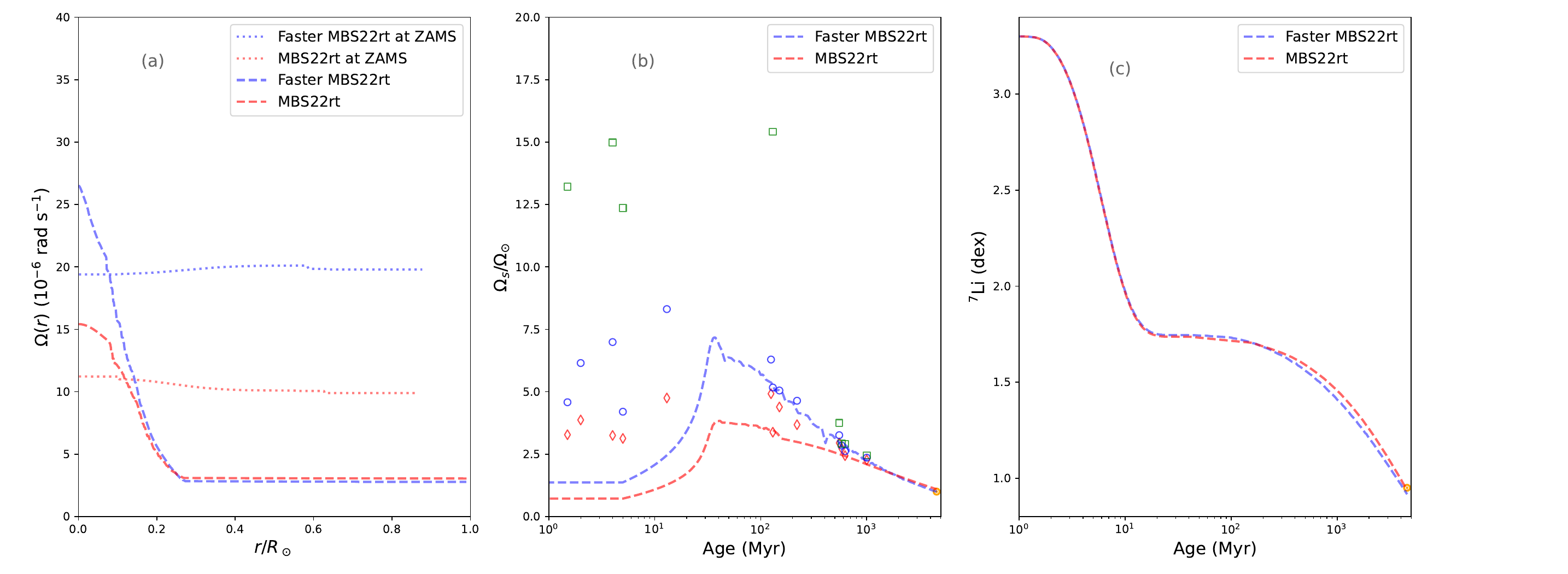}
\caption{(a) Rotation profiles of models at the ZAMS and the age of $4.57$ Gyr.
The age of the solar models at the ZAMS is about $40$ Myr. (b) Surface angular velocity
as a function of age for different solar models. Open symbols refer to observations of surface
angular velocities in open clusters taken from Table 1 of \citet{gall15}, with green, blue, and
red symbols representing the 90th, 50th, and 25th rotational percentiles, respectively.
The orange circle in the lower right indicates the surface rotation rate of the Sun. (c)
Surface lithium abundances as a function of age for different solar models.
\label{fig5}}
\end{figure*}

Moreover, Figure \ref{fig5} shows that MBS22rt predicts a nearly flat rotation profile in the
outer part of the radiative region, with an increase in the rotation rate in the solar core.
These results are consistent with the predictions of models by \citet{yang16} and \citet{egge19,
egge22}, as well as with those inferred from helioseismology \citep{thom03, garc08}. Additionally,
MBS22rt reproduces the surface rotation rates observed in MS solar-type stars in open clusters.

The tachocline affects the surface abundances of Li and Be in the Sun differently due to
the distinct physical properties of these elements; specifically, lithium burns at a lower
temperature than beryllium. The temperature at the base of tachocline in MBS22rt is around
$2.6\times10^{6}$ K, which is high enough to destroy Li but not enough to effectively burn Be.
The existence of a tachocline makes Li in the CZ more easily depleted. Due to the low mixing
efficiency at the base of the tachocline in a low-efficiency rotating model, Be in
the tachocline cannot be quickly transported to a higher-temperature region to be burned. Rotational
mixing below the CZ partially counteracts the gravitational settling of Be within and below the CZ.
Therefore, the presence of the tachocline impedes the depletion of Be but accelerates the depletion of Li.
Accurately and precisely determining the Sun's Be abundance will provide valuable insights into the Sun's
interiors and the physical processes occurring below the CZ.

\begin{deluxetable*} {lcccccccc}
\tablecaption{Measured and Predicted Solar Neutrino Fluxes.
\label{tab2}}
\tablehead{
\colhead{Model} & \colhead{$pp$} & \colhead{$pep$} &\colhead{$hep$} &\colhead{$^{7}$Be} & \colhead{$^{8}$B} & \colhead{$^{13}$N} & \colhead{$^{15}$O} &\colhead{$^{17}$F}  }
\startdata
 Measured &  6.06$^{+0.02 a}_{-0.06}$ & 1.6$\pm$0.3$^{b}$ &...& 4.84$\pm$0.24$^{a}$& 5.21$\pm$0.27$^{c}$ &...& ...&...\\
 B16$^{d}$ &  5.97$^{+0.04}_{-0.03}$ & 1.448$\pm$0.013 & 19$^{+12}_{-9}$ & 4.80$^{+0.24}_{-0.22}$ & 5.16$^{+0.13}_{-0.09}$ & $\leq$13.7 & $\leq$2.8 & $\leq$85 \\
Borexino$^{e}$ & 6.1$\pm$0.5 & 1.39$\pm$0.19 & $<$220 & 4.99$\pm$0.11 & 5.68$^{+0.39}_{-0.41}$ & \multicolumn{2}{c}{6.7$^{+1.2}_{-0.8}$} &  \\
\hline
 MBS22   &  5.99 & 1.446 & 9.75 & 4.85 & 5.25 & 4.02 & 2.00 & 4.86  \\
 MBS22r1 &  5.98 & 1.445 & 9.67 & 4.97 & 5.55 & 4.23 & 2.12 & 5.16  \\
 MBS22r2 &  5.99 & 1.446 & 9.67 & 4.97 & 5.55 & 4.23 & 2.12 & 5.16 \\
 MBS22r3 &  5.98 & 1.442 & 9.67 & 4.97 & 5.55 & 4.21 & 2.11 & 5.14 \\
 MBS22rt &  5.98 & 1.444 & 9.67 & 4.97 & 5.55 & 4.23 & 2.12 & 5.16 \\
\enddata
\tablenotetext{}{Notes. The fluxes of $pp$, $pep$, $hep$, $^{7}$Be, $^{8}$B, $^{13}$N, $^{15}$O, and $^{17}$F neutrinos are in units
of $10^{10}$, $10^{8}$, $10^{3}$, $10^{9}$, $10^{6}$, $10^{8}$, $10^{8}$, and $10^{6}$ $\mathrm{cm}^{-2}\ \mathrm{s}^{-1}$, respectively.}
\tablenotetext{a}{\citet{bell11}.}
\tablenotetext{b}{\citet{bell12}.}
\tablenotetext{c}{\citet{ahme04}.}
\tablenotetext{d}{\citet{berg16}.}
\tablenotetext{e}{\citet{bore18}. The total fluxes, $\Phi(\rm CNO)$, produced by CNO cycle are given by \citet{basi23}.}
\end{deluxetable*}

Moreover, the fluxes of $pp$, $pep$, $hep$, $^{7}$Be, and $^{8}$B neutrinos, as well as the total fluxes of
$^{13}$N, $^{15}$O, and $^{17}$F neutrinos computed from the rotating model MBS22rt, agree with those reported by
\citet{berg16}, \citet{bore18}, \citet{appe22}, and \citet{basi23} at the level of $1\sigma$ (see Table \ref{tab2}).
This indicates that the temperature and density distributions in the nuclear reaction region of the model align
with those of the Sun.

\section{Discussion and Summary}

\subsection{Discussion}
The SSM MBS22 was constructed using OP opacity tables. SSMs constructed with OPAL opacity
tables are not as good as those constructed with OP opacity tables \citep{yang24}. To reproduce
the seismically inferred sound speed and density profile, a linearly increased OPAL opacity and
enhanced diffusion, as suggested by \citet{yang24}, are still necessary. The OPAL opacities
used in the rotating models were linearly increased following \citet{yang24}. Enhanced diffusion
results in a higher initial metal abundance and more metals in the radiative region, while leading to
a lower metal abundance in the CZ. These effects are similar to those of mass accretion with a variable $Z$
before the ZAMS \citep{kuni21, kuni22}, which may deserve further detailed study in future work, along with
the impact of increased opacity. Nevertheless, calculations show that the predicted Li and Be abundances,
as well as the rotation profile, are primarily determined by convective overshoot and rotational mixing,
including magnetic fields, rather than by these factors.

The isotope $^{6}$Li was included in this work, but $^{10}$Be was not considered. In our calculations,
$^{6}$Li was completely burned out before the ZAMS, which contradicts observation, as $^{6}$Li has been
detected on the solar surface. Additional mechanisms are necessary to explain the presence of $^{6}$Li.

Fast rotators in NGC 2264 \citep{bouv16} and the Pleiades \citep{bouv18} have been found to be
systematically richer in Li than their slowly rotating counterparts. \citet{egge12} found that stars
with longer disk-locking periods tend to be more Li depleted, rotate more slowly, and exhibit lower Li
abundances at the ZAMS. The timescale of disk-locking of the Sun during the early phase of pre-MS remains unknown.

\begin{figure}
\includegraphics[angle=0, scale=0.5]{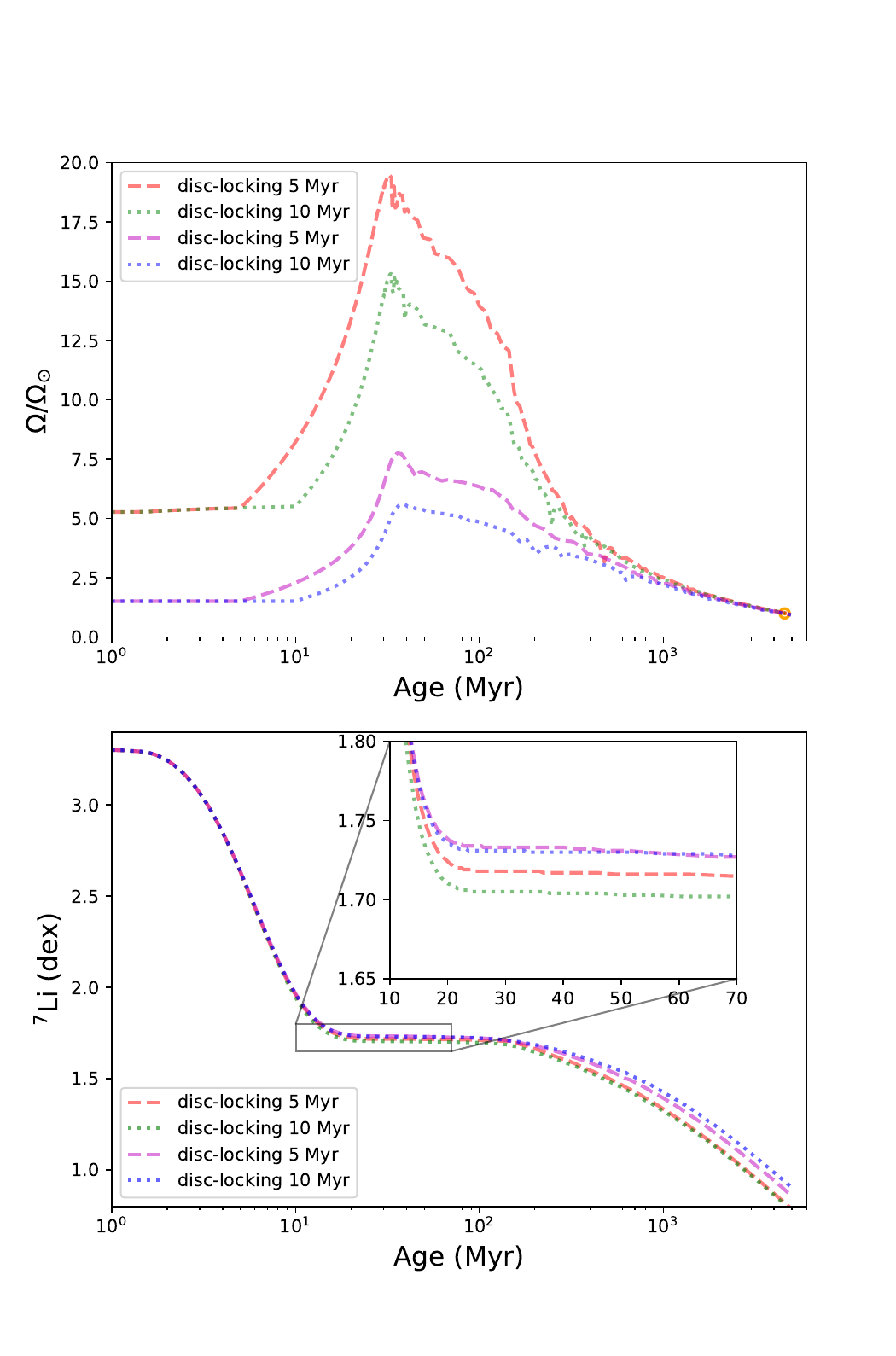}
\caption{(a) Evolution of the surface angular velocity for 1 \dsm{} models with the same initial conditions
except for initial velocities and disk lifetimes. These models are not calibrated to the Sun.
(b) Evolution of the surface lithium abundance of these models.
\label{fig6}}
\end{figure}

In our calculations, we assumed a timescale of 5 Myr. For a given disk-locking timescale, models with
higher $\Omega_{i}$ exhibit higher rotation rates and lower Li abundances at the ZAMS than models with
lower $\Omega_{i}$ (see Figure \ref{fig6}). For a given initial rotation rate $\Omega_{i}$, models with
a shorter disk lifetime tend to have higher Li abundances and rotation rates at the ZAMS compared to those
with longer disk-locking phases. These results are consistent with those reported by \citet[see their
Figures 5 and 8]{egge12} and \citet[see their Figure 4.6]{marq13}. For a given low initial rotation rate,
the timescale does not significantly affect our results (see Figure \ref{fig6}).

From the birth of a star to its ZAMS, rotating models that include magnetic field effects behave
almost like homogenous solid-body rotation models (see Figure \ref{fig5}), except for the distributions of Li and Be,
which have been burned out in the center of the models (see Figure \ref{fig3}). The surface rotation rate
reaches its maximum value around the ZAMS due to contraction, so we take this value, $\Omega_{\rm z}$,
to characterize the initial rotation rate.

The initial rotation rate within a certain range does not significantly affect our results. For example,
the rotating model with $\Omega_{\rm z}\simeq 20\times10^{-6}$ rad s$^{-1}$ can also reproduce the surface
He, Li, and Be abundances of the Sun. However, this model has a faster rotating core than MBS22rt (see Figure \ref{fig5}),
which appears inconsistent with helioseismic results. When the initial rotation rate $\Omega_{\rm z}$
reaches approximately $30\times10^{-6}$ rad s$^{-1}$ ($\approx 11\ \Omega_{\odot}$), the model predicts
not only a surface Li abundance lower than that advocated by \citet{wang21} but also an excessively
fast core rotation. This occurs because the angular momentum stored in the Sun's radiative region must be transferred
to the CZ through the BCZ and then lost through the CZ. The more angular momentum stored at the ZAMS, the more
intense the mixing at the BCZ during the MS, leading to excessive Li depletion, consistent with
the calculations of \citet{marq13}. Consequently, both helioseismic results and the observed Li abundance seem to
favor an initial solar rotation rate with $\Omega_{\rm z} < 10\ \Omega_{\odot}$, suggesting that the Sun might
have been a slow rotator at the ZAMS rather than a fast rotator. However, this model does not align with observations
of the Pleiades \citep{bouv18}, where faster rotators exhibit less Li depletion.

The rotation rate $\Omega_{\rm z}$ depends on the initial rotation rate, disk-locking timescale, magnetic
braking, and internal angular momentum transport. The evolution of surface rotation rate during the MS phase is
primarily determined by the torque acting on the Sun and internal angular momentum transport. Although our models
can reproduce the evolution of the surface rotation rates speculated from solar-type MS stars in clusters, the initial
rotation rate during the pre-MS phase is lower than expected (see Figure \ref{fig5}).

In this work, we considered only the angular momentum loss mechanism of \citet{kawa88}. It can reproduce
the solar rotation rate at an age of 4.57 Gyr and the evolution of the surface rotation rates inferred from
solar-type MS stars in clusters (see Figure \ref{fig5}). In addition to the \citet{kawa88} relation,
other magnetic braking laws, such as those of \citet{rein12} and \citet{matt12, matt15}, exist.
These laws prescribe angular momentum loss rates that follow different power laws, which could affect
the predicted Li and Be abundances. Therefore, Li and Be abundances may provide an opportunity
to test these different formulas and the initial conditions of the Sun.

In our calculations, the value of $f_{c}$ is 0.03. This is approximately consistent with the value of 0.046
adopted by \citet{pins89} and the 0.023 reported by \citet{hunt08} but it is significantly lower than
the 0.8 suggested by \citet{prat16}.

\begin{figure}
\includegraphics[angle=0, scale=0.5]{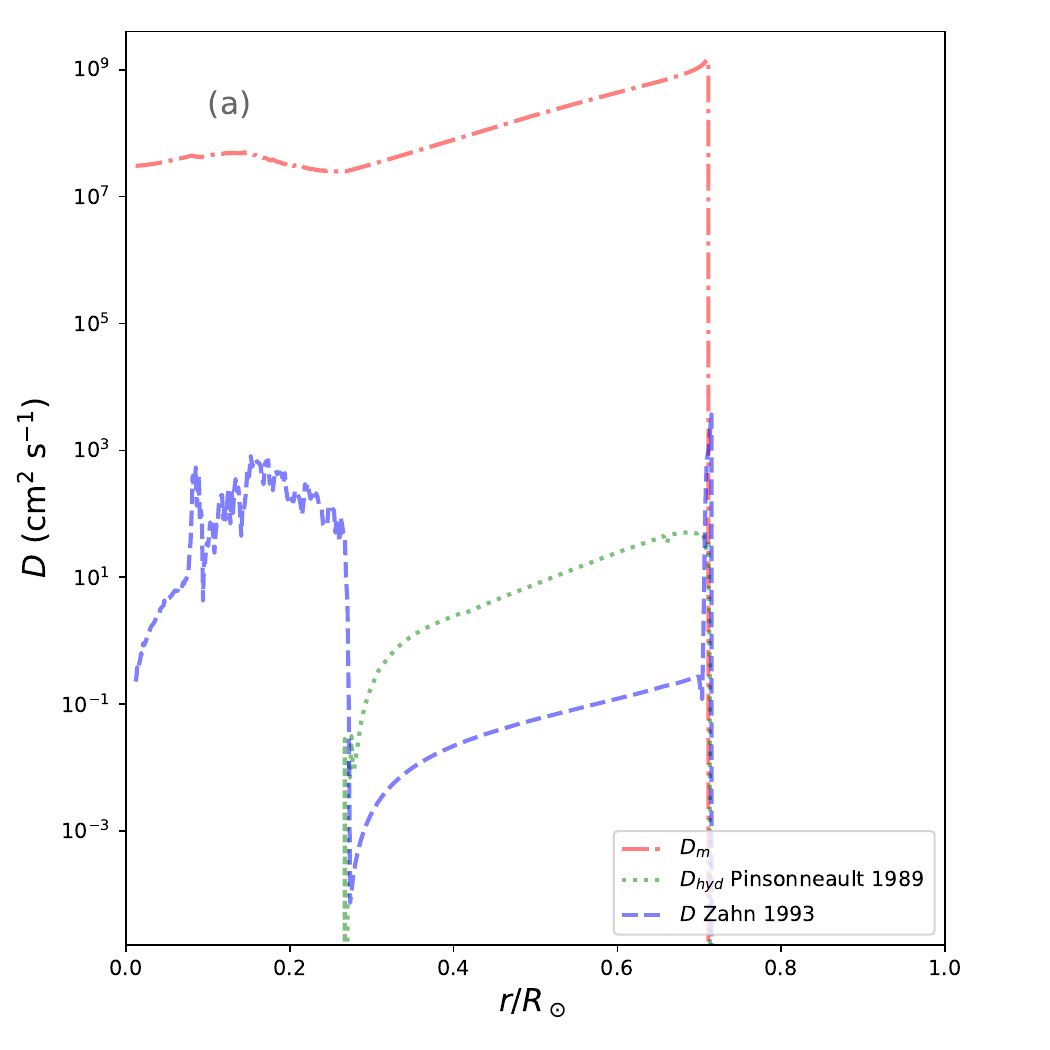}
\caption{Hydrodynamic and magnetic diffusion coefficients as a function of radius for MBS22rt.
\label{fig7}}
\end{figure}

Angular momentum transport and material mixing in rotating models are dominated by magnetic field effects.
The magnetic diffusion coefficient is much larger than the hydrodynamic diffusion coefficients (see
Figure \ref{fig7}). Solar models without magnetic fields cannot replicate the inferred nearly flat rotation
profile in the radiative region \citep{egge22, yang24}. The value of $f_{m}$ is determined by both the inferred
flat rotation profile and the observed Li and Be abundances. A slight decrease or increase in $f_{m}$
necessitates a corresponding increase or decrease in $f_{cm}$ to match the observed Li and Be abundances.

\citet{spru02} and \citet{maed05} demonstrated that, when $\omega_{A} \ll \Omega$, where $\omega_{A}$
is the Alfvén frequency, the coefficient for chemical transport is reduced by a factor $(\omega_{A}/\Omega)^{2}$
compared to that for angular momentum transport. In our calculations, the parameter $f_{cm}$ is assumed to be
constant. The value of $\omega_{A}/\Omega$ varies with radius and model age, ranging from approximately
$10^{-2}$ to $10^{-4}$, and is about $10^{-2}$ at the BCZ.

The mixing efficiency in the CZ can also influence Li and Be abundances predicted by models. In our calculations,
we determine the amounts of Li and Be consumed in each shell and assume that the material within the CZ is fully
mixed instantaneously at each evolutionary time step.

Based on multidimensional simulations, \citet{bara17} proposed a new expression for the diffusion coefficient
of overshooting. They assumed a maximum penetration depth, $d_{\rm ov}$, that depends on the rotation rate, with fast rotation
strongly limiting the vertical penetration of the convective plumes. Specifically, they set $d_{\rm ov}$ = 1 $H_{p}$
if $\Omega <$ 5 $\Omega_{\odot}$ and 0.1 $H_{p}$ otherwise. Convective overshooting plays an important role in depleting
Li during the pre-MS stage, as a larger overshooting distance results in a higher Li-burning rate in the overshooting region.
\citet{augu19} later modified this formula. In this work, we adopted the simplest classical treatment for convective overshoot.

Rotationally dependent overshooting has been used to explain the observed correlation between rotation and Li depletion
in young clusters \citep{bara17, dumo21a, cons21}. \citet{dumo21a} showed that, in addition to atomic diffusion,
meridional circulation, and turbulent shear, additional parametric turbulent mixing processes are required to simultaneously
explain the observed Li depletion and the solar rotational profile. However, this parametric turbulent mixing leads to a 0.3 dex
depletion of Be by the age of the Sun, which is slightly too large compared to observations.

\citet{cons21} demonstrated that fast rotation can suppress convection, thereby decreasing the temperature
at the BCZ to an extent sufficient to account for the Li spread observed in young open clusters. The suppression
effect is proportional to $\Omega^{2}$ and, therefore, is not expected to be significant for slow rotators.
However, \citet{cons21} did not consider microscopic diffusion, the reduction of effective gravity due to centrifugal
force, or rotational mixing. If the suppression effect exists in the Sun, reproducing the seismically inferred convection
depth would be more challenging. In this work, rotationally dependent overshooting and the suppression effect
are not included. More studies are needed to fully understand the behavior of Li and Be in star clusters.

The settling of heavy elements has been confirmed by helioseismology, and thus, Be must undergo
the gravitational settling. Only a portion of the Be deposited in the tachocline can be counteracted by
rotational mixing, while most of it is transported to higher-temperature regions by settling, diffusion,
and rotational mixing, where it is burned. Thus, the Be abundance predicted by a rotating model cannot be
significantly higher than that predicted by the SSM MBS22. The maximum value of the surface Be abundance predicted
by rotating models is $1.34$ dex. This implies that \citet{aspl21} might have overestimated the solar surface Be
abundance.

\citet{pins89} and \citet{egge22} have shown that rotational mixing contributes to Li depletion in the Sun.
The models of \citet{yang24} did not include the gravitational settling of Li and Be or the effect of tachocline,
predicting a Li abundance higher than observed. Both \citet{egge22} and our study considered the effects of
magnetic fields. The rotation profiles in our models are similar to those of \citet{egge22}. However, their models
predict a surface Be abundance lower than the observed value. If the gravitational settling of Li and Be is neglected
in our calculations, our models can also reproduce the observed Li abundance but predict a Be abundance lower than observed.
Neglecting the effect of the tachocline exacerbates these discrepancies. Additionally, the impact of accretion with
varying abundances, which could explain the difference between Li and Be, is not considered in this work. The Li
abundance predicted by the accretion models of \citet{zhang19} is lower than observed, and they did not
consider Be.

\subsection{Summary}
In this work, we constructed standard and rotating solar models based on Magg’s mixtures.
The surface helium abundance and the CZ depth of the SSM do not match the seismically inferred values.
The surface Li abundance predicted by the SSM is inconsistent with the observed value.

To reproduce the helioseismic results and the observed Li abundance, the effects of rotation, magnetic
fields, and gravitational settling must be considered. Consequently, we developed the rotating model, MBS22rt,
which outperforms the SSM and the earlier rotating models, such as those in \citet{egge22} and \citet{yang24}.
It assumes convective overshooting with $\delta_{\rm ov}=0.09$ and a tachocline width of $0.05R$.
MBS22rt exhibits superior sound speed and density proﬁles compared to the SSM and successfully reproduces
the observed ratios $r_{02}$ and $r_{13}$. The surface helium abundance and the CZ depth of MBS22rt are consistent
with seismically inferred values within $1 \sigma$. Furthermore, the fluxes of $pp$, $pep$, $hep$,
$^{7}$Be, and $^{8}$B neutrinos, as well as the total fluxes of $^{13}$N, $^{15}$O, and $^{17}$F
neutrinos, calculated from MBS22rt, agree with those reported by \citet{berg16}, \citet{bore18}, and \citet{basi23}
at the level of $1\sigma$.

Moreover, the rotating model predicts both a nearly ﬂat rotation proﬁle in the external part of the radiative
region and an increase in the rotation rate in the solar core, which is in good agreement with that
inferred from helioseismology \citep{thom03, garc08}. The surface Li abundance of 0.94 dex predicted
by MBS22rt agrees with the value of 0.96 $\pm$ 0.05 dex reported by \citet{wang21}, while the surface Be
abundance of 1.31 dex is consistent with 1.32 $\pm$ 0.05 dex reported by \citet{koro22}. The rotating
model can simultaneously reproduce the Sun's He, Li, and Be abundances, the seismically inferred CZ depth, the
rotation profile, and other helioseismic results. This indicates that rotation, including magnetic fields,
plays a key role in the evolution of the Sun. Since Li and Be have different
physical properties, their depletion is governed by different physical mechanisms.
The depletion of Be in the CZ of the Sun is dominated by gravitational settling but is affected by the
tachocline and rotational mixing, whereas the depletion of Li in the CZ is mainly determined by convective
overshooting and rotational mixing. The presence of the tachocline and rotational mixing accelerates
the depletion of Li abundance in the Sun but slows down Be depletion. These distinct depletion
mechanisms result in the surface Li and Be evolving differently over time. The depletion of Li
and Be also suggests that the helium abundance on the surface of the Sun has been enhanced by rotational mixing.
An SSM with the correct metal abundance would underestimate the surface helium abundance.

\begin{acknowledgments}
Authors thank the anonymous referee for many useful suggestions that helped the authors signiﬁcantly improve this work,
as well as J. W. Ferguson for providing their low-temperature opacity tables, and acknowledge the support from the NSFC 11773005,
12222301, and the CSST Project. TZJ acknowledges the support from NSFC 11803030 and Yunnan Province Basic Research Program
Project 202101AT070021.
\end{acknowledgments}

\end{document}